# The rise of the chaebol: A bibliometric analysis of business groups in South Korea

Artur F. Tomeczek

SGH Warsaw School of Economics

artur.tomeczek@sgh.waw.pl

**Abstract:** South Korea has become one of the most important economies in Asia. The largest Korean multinational firms are affiliated with influential family-owned business groups known as the chaebol. Despite the surging academic popularity of the chaebol, there is a considerable knowledge gap in the bibliometric analysis of business groups in Korea. In an attempt to fill this gap, the article aims to provide a systematic review of the chaebol and the role that business groups have played in the economy of Korea. Three distinct bibliometric networks are analyzed, namely the scientific collaboration network, bibliographic coupling network, and keyword co-occurrence network.

# 1. Introduction

South Korea has become one of the largest and most important economies in Asia. Korean multinational firms are a major source of innovation and investment flows, both regionally and globally. Korea's prominent family-owned business groups are known as the chaebol. In the last decades, the chaebol have attained the spotlight of economic sciences, much like the Japanese keiretsu networks had done since the 1980s (Tomeczek, 2022).

Despite the surging academic popularity of the chaebol, there is a considerable knowledge gap in the bibliometric analysis of business groups in Korea. This article attempts to fill this gap. Bibliometrics is currently one of the most popular ways of conducting empirical studies of scientific collaboration as well as systematizing and mapping existing knowledge. Such studies can be based on quantitative data representing collaborations between scientists or institutions (co-authorship networks), keywords assigned to scientific articles (keyword co-occurrence networks), and scientific citations (citation networks, bibliographic coupling networks, and co-citation networks).

Using the bibliometric methodology, this article aims to provide a systematic review of the chaebol and the role that business groups have played in the economy of Korea. To fulfill this aim, three research questions are stated. [RQ1] Which connections are the most influential in the scientific collaboration network? [RQ2] How many communities are there in the bibliographic coupling network and what are their characteristics? [RQ3] What are the main themes of analysis in the keyword co-occurrence network? The answers to these questions are relevant to the systematization of academic research on chaebol groups.

This article is divided into six sections. The research methodology is detailed in section two. Section three provides a broad historical background to chaebol groups and the economic development of Korea. Section four explores the results of the bibliometric analysis. Implications of the research are discussed in section five. The final section concludes the article.

## 2. Methodology

*Bibliometric networks*

The origins of bibliometric and scientometric analysis can be traced to the seminal works of Garfield (1955), Kessler (1963a, 1963b), Price (1965), and Small (1973). Over the decades, and with the evolution of digital technologies, bibliometrics has gained numerous tools that provide new and exciting opportunities for researchers.

To answer the first research question, a co-authorship network is constructed based on the data aggregated to the level of country/territory. The network consists of nodes (vertices) representing countries/territories and edges (links) representing the number of co-authored articles. Co-authorship networks, at different levels of aggregation, are likely the most common type of network in bibliometric studies (Abbasi, Hossain, & Leydesdorff, 2012; Barabási et al., 2002; Beaver & Rosen, 1978; Gazni & Didegah, 2011; Glänzel, 2001; Glänzel & Schubert, 2001, 2005; Newman, 2001; Uddin, Hossain, Abbasi, & Rasmussen, 2012; Yan, Ding, & Zhu, 2010).

Bibliographic coupling provides the answer to the second research question. In this particular network, nodes are articles and edges connecting them are sources referenced in both articles. Bibliographic coupling networks can be used to identify clusters of publications that are based on the same underlying concepts (Chen, Zhang, & Fu, 2019; Fernandes & Ferreira, 2022; Yadav, Kumar, & Malik, 2022; R. Zhang & Yuan, 2022). The results of bibliographic coupling analyses tend to be relatively similar to co-citation networks (Yan & Ding, 2012).

The main themes of analysis, explored in the third research question, are identified using a keyword co-occurrence network. Nodes represent keywords and edges represent the fact that they are assigned to the same article. Keyword co-occurrence networks are a popular way to conduct thematic knowledge mapping of the existing body of literature (P.-C. Lee & Su, 2010; Oliveira, Carvalho, & Reis, 2022; Su & Lee, 2010; Szczech-Pietkiewicz, Radło, & Tomeczek, 2023; Wang, Huangfu, Dong, & Dong, 2022).

*Data collection and previous studies*

A systematic and transparent data collection process is of paramount importance for bibliometric analysis. Scopus, a popular database, is used to collect bibliometric data. Following numerous bibliometric studies on various topics (Fernandes & Ferreira, 2022; Jiang, Ritchie, & Benckendorff, 2019; P.-C. Lee & Su, 2010; Li et al., 2021; Su & Lee, 2010; Velez-Estevez, García-Sánchez, Moral-Munoz, & Cobo, 2022; Wang et al., 2022; Xu, Hou, & Wang, 2022; Yadav et al., 2022; X. Zhang, Xie, Song, & Song, 2022), the publications analyzed in this study are limited to peer-reviewed scientific articles published in English. Furthermore, the number of articles is limited to include at least one of the four subject areas relevant to business groups (Business, Management, and Accounting; Economics, Econometrics, and Finance; Social Sciences; Decision Sciences). To improve the reproducibility of the results, articles published after 2022 are ignored. The detailed search query string created for this article is provided in TABLE 1. It can be used in the "advanced search" in Scopus without any modifications. The search results are 749 articles (this includes both research articles and review articles). However, due to missing information on some articles, such as author keywords, the actual number differs depending on the need of the network (see: TABLE 1). This article also includes references to publications beyond the 749 chaebol-related articles identified in Scopus (i.e., bibliometric methodology, comparison to Japanese keiretsu networks, and the general history of Korea), but those publications are excluded from the bibliometric analysis. The list of 749 articles is available in the associated dataset (Tomeczek, 2023).

As per the results of the search in the Scopus database, this article is the first study of its kind, utilizing bibliometric methodology in the analysis of the Korean chaebol groups. There are, however, previous bibliometric and scientometric studies based on Korean data focusing on other issues such as time allocation of academics (J.-K. Jung & Choi, 2022), collaborations between physicists (M.-J. Kim, 2001), collaborations with Chinese researchers (H. W. Park, Yoon, & Leydesdorff, 2016), online presence of scientific journals (Holmberg & Park, 2018), and knowledge creation and the triple helix model of innovation systems (H. Choe & Lee, 2017; Choung & Hwang, 2013; K.-S. Kwon, Park, So, & Leydesdorff, 2012; J. Yoon, 2015).

TABLE 1 Data collection process

| Search query string (database: Scopus, date: February 2, 2023) | Results |
|---|---|
| TITLE-ABS-KEY ( "chaebol*" OR "jaebeol*" ) OR TITLE-ABS-KEY ( "business group" OR "business network" OR "corporate group" OR "corporate network" OR "conglomerate" OR "holding company" AND "korea*" ) AND PUBYEAR < 2023 AND DOCTYPE ( ar OR re ) AND LANGUAGE ( english ) AND SUBJAREA ( busi OR deci OR econ OR soci ) | 749 |
| Articles with references data (for bibliographic coupling) | 719 |
| Articles with affiliations data (for co-authorship) | 705 |
| Articles with author keywords data (for keyword co-occurrence) | 583 |

Source: Own elaboration based on Scopus (Elsevier, 2023).

*Data preparation and software*

Modularity analysis is a popular method in the study of bibliometric networks. Its purpose is the identification of communities of nodes. This method can be applied to all three types of networks proposed in this article. Common examples include the clusters in international collaboration networks (showing the regionality in co-authorship), bibliographic coupling and co-citation networks (showing the influence of the authors), and keyword co-occurrence networks (showing common combinations of keywords).

Keyword standardization is a required step in data preparation for co-occurrence networks. Including closely related keywords as separate nodes in the network can distort the results. For this article, normalizations include synonymous keywords (e.g., "chaebol" and "korean business groups"), abbreviations (e.g., "foreign direct investment" and "fdi"), spellings (e.g., "globalization" and "globalisation"), and plural and singular forms (e.g., "family firms" and "family firm"). Additionally, some Korea-specific keywords are aggregated to their generic version, since all the analyzed articles are about Korea (e.g., "smes" and "korean smes"). Importantly, when a group of keywords is aggregated, the keyword with the highest occurrence count in that group is chosen as the name of the group. A list of all keywords (raw and aggregated) is provided in the associated dataset (Tomeczek, 2023).

The data downloaded from Scopus are imported into VOSviewer, a software for bibliometric analysis (van Eck & Waltman, 2010). The networks are then exported to Gephi which is used for network analysis and visualizations (Bastian, Heymann, & Jacomy, 2009). Gephi uses the algorithm of Blondel et al. (2008) for modularity calculations. In modularity analysis, resolution determines the size of the community. Gephi's modularity resolution is based on Lambiotte et al. (2014). Throughout this article normalized citation counts are used to facilitate comparison between newer and older articles. They are defined as "the number of citations of the document divided by the average number of citations of all documents published in the same year and included in the data that is provided to VOSviewer" (van Eck & Waltman, 2022, p. 38). Detailed results and high-quality graphs are included in the associated dataset (Tomeczek, 2023).

## 3. Overview of chaebol groups

*Historical background and comparison to keiretsu networks*

Korea's high rate of economic growth and rapid development caused a deserved spike in scientific interest. After the 1963 elections, the Park Chung-hee government's interventionist policies strongly reinforced the expansion of Korean exports (Haggard, Kim, & Moon, 1991; K.-Y. Jeong & Masson, 1990). For Korea, intra-industry trade is an important channel of business cycle synchronization with other Asian economies (K. Shin & Wang, 2004). Korean investment in China is geographically clustered and has a positive relationship with regional income, labor quality, low wages, the number of economic zones, and proximity to Korea (S. J. Kang & Lee, 2007). The negative interest rate policy of Japan caused an increase in stock prices in emerging Asian economies, but this effect was absent specifically for the Korean stock market (Fukuda, 2018). Human capital growth is stable and significantly contributes to the economic growth of Korea, despite the aging workforce (J.-S. Han & Lee, 2020). Haggard (2018) provides a comprehensive review of the leading theories explaining the economic growth of the East Asian developmental states.

In its rapid transition to an advanced economy, Korea has in many ways mirrored Japan's earlier economic miracle. Likewise, Korean chaebol groups are often compared to Japanese keiretsu networks, as both of them have garnered substantial interest from domestic and international scholars (Tomeczek, 2022). Much like in a keiretsu, firms affiliated with a chaebol form a network of cross-shareholdings. However, there are some important differences between the two – chiefly the role played by the banks. The chaebol have a more hierarchical structure. One of the key challenges faced by business groups in post-war Japan was the fact that holding companies were made illegal (Buckley, 2004; Kōsai, 1989). This created a substantial power vacuum in the market as holding companies were the core of the pre-war zaibatsu groups. After the American occupation of Japan had ended, this vacuum was quickly filled by Japanese commercial banks as they supplied capital and leadership to the scattered firms of the former zaibatsu (Tomeczek, 2022). Inherently, keiretsu networks formed around the commercial banks, known as the main banks, at their center. The primary functions of the main banks are the provision of capital and monitoring the performance of firms through formal and informal ties (Aoki, Patrick, & Sheard, 1994; Sheard, 1994). In general, the chaebol have been more influential in Korea than the keiretsu have been in Japan (Campbell & Keys, 2002).

*Internal markets and corporate governance*

Due to restrictions on commercial bank ownership in Korea, business groups are characterized by an internal financing system (Almeida, Kim, & Kim, 2015; Hicheon Kim, Hoskisson, Tihanyi, & Hong, 2004; S. Park & Yuhn, 2012). Investments in non-bank financial firms by chaebol groups were commonplace (Hicheon Kim et al., 2004; Jiyoung Kim, 2017). Business groups have been shown to engage in the propping (negative tunneling) of financially distressed firms (G. S. Bae, Cheon, & Kang, 2008). During the 1970s, the Korean government provided massive financial aid to the largest chaebol groups through state-owned banks, at the expense of small and medium enterprises (Haggard & Moon, 1990).

The ownership of chaebol groups is much more centralized and hierarchical, and their central planning offices (holding companies) are still controlled by influential families (Almeida, Park, Subrahmanyam, & Wolfenzon, 2011; G. S. Bae et al., 2008; Byun, Choi, Hwang, & Kim, 2013; H. C. Kang, Anderson, Eom, & Kang, 2017; E. Kim, 2006; Oh & Park, 2001; C.-K. Park, Lee, & Jeon, 2020; S. Park & Yuhn, 2012). In that sense, the chaebol are more similar to the pre-war zaibatsu groups than the modern keiretsu networks. Acquisitions of new firms that are placed into the pyramidal structures of chaebol groups have a negative impact on acquiring firms' stock prices (Almeida et al., 2011; K.-H. Bae, Kang, & Kim, 2002). As such, if the market expects future acquisitions by a chaebol firm to be at the expense of its minority shareholders, the stock of that firm will be undervalued. Korean companies usually have a lower price-earnings ratio (Ducret & Isakov, 2020). The relations between affiliated firms are complex and the controlling families do engage in intragroup tunneling (K.-H. Bae et al., 2002; Baek, Kang, & Lee, 2006; Oh, Yoon, & Kim, 2022) and propping (G. S. Bae et al., 2008). Alternatively, instead of tunneling, the position of the firms in the pyramid can be explained by their profitability before the acquisition, as highly-profitable firms are more likely to be placed under the direct control of the family (Almeida et al., 2011). Family ownership concentration has a positive impact on chaebol firms' productivity, but non-chaebol firms tend to have higher productivity than chaebol firms (E. Kim, 2006). Internal transactions of chaebol groups can cause market distortions and deter potential entrants (Jin, 2020).

Executive compensation is less likely to be performance-related in chaebol firms compared to non-chaebol firms (Kato, Kim, & Lee, 2007), but the chance increases when a Korean firm has foreign investors (Garner & Kim, 2013). Foreign ownership has a positive impact on both innovation (Joe, Oh, & Yoo, 2019) and, specifically for chaebol-affiliated firms, bankruptcy likelihood (Jounghyeon Kim, 2020). The compensation of chief executives who are family members is generally substantially higher (Hohyun Kim & Han, 2018). Additionally, top-executive turnover in the largest chaebol groups is not tied to performance (Campbell & Keys, 2002). Intragroup executive transfers are common and many chief executives retain influence over their former companies (C. Kim, Park, Kim, & Lee, 2022).

*Asian financial crisis and market liberalization*

The 1997 Asian financial crisis simultaneously limited the scale of internal market financing in chaebol groups (S. Lee, Park, & Shin, 2009) and caused a spike in their investments in non-bank financial firms (Jiyoung Kim, 2017). The existence of internal capital markets of the groups was beneficial to them during the crisis period (Almeida et al., 2015). Before the crisis, Korea experienced an investment boom focusing on the manufacturing sector which was financed primarily by extensive short-term corporate debt (Haggard & Mo, 2000). Despite heavy pressure to liberalize the economy, many of the developmental state policies have persisted after the financial crisis (Dalton & Rama, 2016; Ha & Lee, 2007; I. Jun, Sheldon, & Rhee, 2010; Y. S. Park, 2011). The crisis sparked many crucial changes in Korea, such as improvements in corporate transparency (J. Chang, Cho, & Shin, 2007; S. Choe & Pattnaik, 2007), the evolution of the general corporate culture (Cho, Yu, Joo, & Rowley, 2014), an increase in inward FDI (Fitzgerald & Kang, 2022), and the expansion of welfare state (H. K. Song, 2003). This period has also seen significant tax reforms (S. H. Park, 2022) and the restructuring of the service sector (S. I. Shin & Kim, 2020).

The chaebol have a non-linear (inverted U) impact on industry innovation (C.-Y. Lee, Lee, & Gaur, 2017; Mahmood & Mitchell, 2004). In this context, business groups are highly beneficial to technological catch-up when institutions are weak and technology has low appropriability (S. J. Chang, Chung, & Mahmood, 2006; C.-Y. Lee et al., 2017; Mahmood & Mitchell, 2004; K.-H. Park & Lee, 2006). Similarly, chaebol affiliation tends to positively influence affiliated firms' labor productivity (B. Jung, Lee, Rhee, & Shin, 2019). On the other hand, non-chaebol firms are shown to be more profitable (Joh, 2003) and have higher efficiency of government research and development grants (I. Kwon & Park, 2021) than chaebol-affiliated firms. Institutional blockholders negatively impact research and development investments (S. Kang, Chung, & Kim, 2019), but they also lower future information asymmetries (Chung, Kim, & Wang, 2022). Domestic blockholders tend to improve corporate governance (K. Y. Lee, Chung, & Morscheck, 2020), unless they induce short-termism (Chung, Kim, & Lee, 2020).

*The political power of the largest chaebol groups*

Historically, the five largest chaebol were Samsung, LG, Daewoo, Hyundai, and SK (Campbell & Keys, 2002; S. Park & Yuhn, 2012). Since 1987, there have been numerous corporate governance reforms aiming to curb the dominant position of the chaebol (Oh et al., 2022). In 1999 the massively mismanaged Daewoo went bankrupt (Joongi Kim, 2008), but the other four continue their market dominance to this day. According to the Annual Report of the Korea Fair Trade Commission (2022, pp. 210–212), there are 71 Korean business groups large enough to be subject to disclosure requirements; 40 of them are subject to cross-shareholding restrictions and five of them have assets exceeding 100 trillion KRW (Samsung, Hyundai Motor, SK, LG, and Lotte).

As a member of the OECD's Development Assistance Committee, Korea is an important donor of official development assistance (E. M. Kim & Oh, 2012). As chaebol groups exert high formal and informal influence on Korean official development assistance, they can use it as a part of their international expansion strategy (Schwak, 2019). The appointment of politically connected outside directors to the board of chaebol firms increases their performance and lowers the risk, at the cost of some monitoring ability (J. Y. Shin, Hyun, Oh, & Yang, 2018). Outside directors have a lower survival chance on the board if they act proactively (T. Yoo & Koh, 2022). The presence of a banker interlocking director on the board of a chaebol-affiliated firm lowers its cost of debt (Nam & An, 2023).

The concentration of money and power increases the risk of corruption. The historical examples of corruption, cronyism, and state capture by influential chaebol groups are infamous and have been documented in numerous studies (Albrecht, Turnbull, Zhang, &

Skousen, 2010; Ha & Lee, 2007; I.-W. Jun, Kim, & Rowley, 2019; B.-K. Kim & Im, 2001; You, 2020; You & Park, 2017). Two recent examples featured in the media include the regulatory capture that likely contributed to the tragic sinking of the Sewol ferry in 2014 (You & Park, 2017) and the impeachment of President Park Geun-hye in 2016 (You, 2020). In general, politics in Korea is highly polarized (Al-Fadhat & Choi, 2023; S. Han, 2022; Oh et al., 2022).

The economic literature remains split on whether the impact of chaebol groups on their affiliated firms is ultimately positive or negative. There is, however, a strong consensus that these large conglomerates have been extremely influential during the rapid economic development of the Korean economy.

## 4.     Results of bibliometric analysis

*Prolific journals, institutions, and scientific collaboration*

Based on the search in the Scopus database, the earliest English-language article was published in 1987 in California Management Review. It is a study of the chaebol management style (S. Yoo & Lee, 1987), that has since garnered 87 citations. GRAPH 1 shows the steady evolution of scientific interest in chaebol groups. Over the decades the number of publications has increased significantly. 2019 was the annual peak with 66 new articles, but 2022 is a close second with 62. The 2020-2021 slowdown is most likely associated with the pandemic-related disruptions. Scientific journals with the highest number of publications are Sustainability (40 articles), Pacific-Basin Finance Journal (24 articles), Asia Pacific Business Review (19 articles), Asia-Pacific Journal of Financial Studies (19 articles), Journal of Contemporary Asia (15 articles), Global Economic Review (14 articles), Emerging Markets Finance and Trade (12 articles), and Journal of Applied Business Research (11 articles).

GRAPH 2 represents the scientific collaboration network. The size of nodes is determined by the number of articles assigned to a country/territory. Edges are scaled according to the co-authorship count. Nodes are only included if they have at least two articles and at least one connection to other nodes. The network is centralized around Korea and its collaborative endeavors with researchers affiliated with the United States (119 articles). Other directions of scientific collaboration are also important but remain significantly less prolific. These include Korea and the United Kingdom (19 articles), Korea and China (10 articles), Korea and Australia (7 articles), Korea and Canada (7 articles), and the United States and Singapore (7 articles).

Unsurprisingly, Korean universities and institutions are the main driving force of academic research into chaebol groups. The most common affiliations are Seoul National University (63 articles), Korea University (52 articles), Korea Advanced Institute of Science and Technology (44 articles), Yonsei University (41 articles), Chung-Ang University (33 articles), Korea University Business School (33 articles), Hanyang University (27 articles), Ewha Womans University (21 articles), Chung-Ang University Business School (20 articles), and KAIST College of Business (20 articles). The most common funding sponsors are the National Research Foundation of Korea (46 articles) and the Ministry of Education (28 articles).

GRAPH 1 The annual number of new scientific articles on chaebol groups

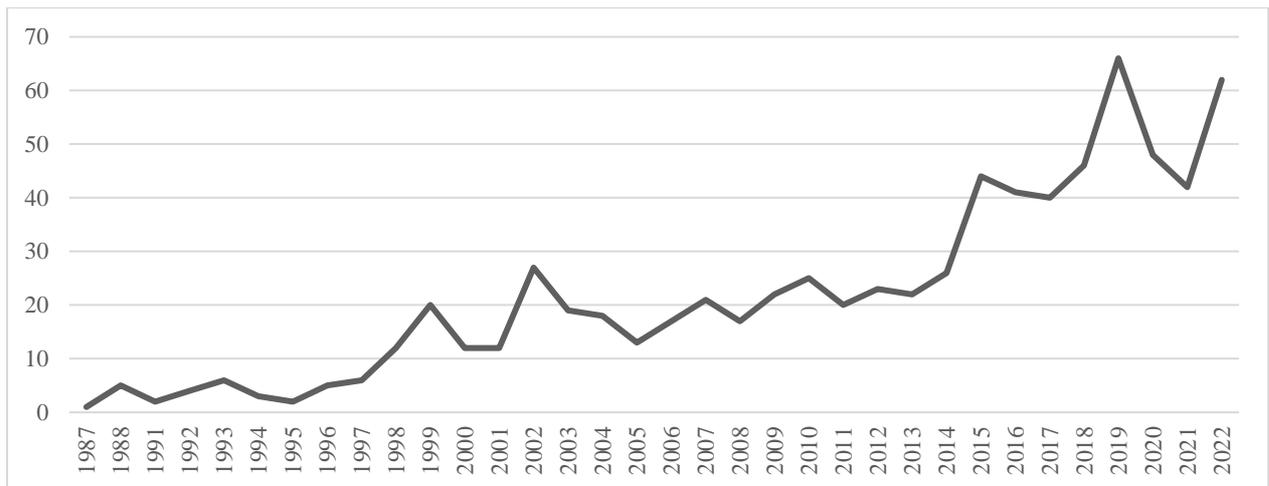

Source: Own elaboration based on Scopus (Elsevier, 2023).

GRAPH 2 Scientific collaboration network (size: number of articles, color: modularity class)

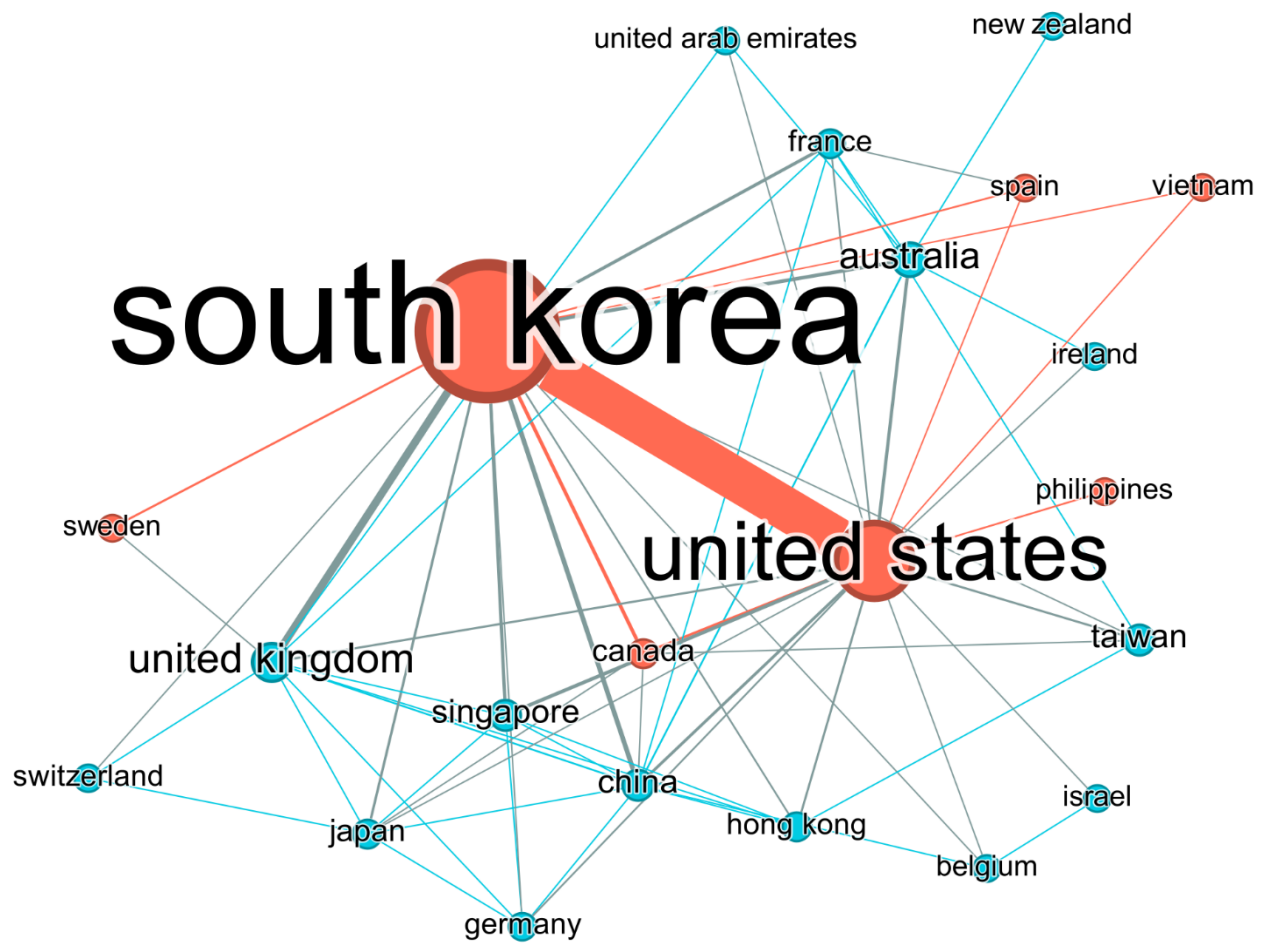

Source: Own elaboration based on Scopus (Elsevier, 2023).

*Bibliographic coupling network and influential articles*

In the bibliographic coupling network, communities are based on the analysis of the edges connecting the articles. If two articles both reference a substantial number of common third articles then it is likely that they are analyzing a similar topic or use comparable methodology. If these communities are distinct enough they might allow researchers to identify influential publications and concepts linking them together. This approach is similar to other reference-based analyses, such as co-citations networks. The following section is an attempt at systematizing the research trends in the existing literature on Korean business groups.

GRAPH 3 and GRAPH 4 visualize the bibliographic coupling networks. The former includes articles with more than 10 citations, while the latter limits that sample to articles with 200 or more citations. Color in GRAPH 3 represents the modularity class (community), while in GRAPH 4 color is determined by normalized citation count. In both networks, node size is determined by citation count and the labels are limited to the first author and year of publication. The results of the bibliographic coupling analysis are summarised in TABLE 2.

Starting with GRAPH 3, the network has 261 nodes (articles). Modularity analysis identifies five communities (resolution = 1.00). The three largest communities (241 articles in total) are highlighted in specific colors in the network (in the order of size: violet, green, and orange), while the two smallest ones are left gray (20 articles in total).

"Economic system," the largest community in the network, has 102 articles. Because of its size, the community includes articles on many aspects of the economic system of Korea. The primary theme of this group of articles is the developmental state – the role that the strong state played in the industrialization and rapid economic development of post-war Korea – and its eventual decline (Amsden & Hikino, 1994; Cherry, 2003; Hundt, 2014; Y. T. Kim, 1999, 2005; Kyung-Sup, 1999; S. J. Lee, 2008; S.-J. Lee & Han, 2006; H. Lim, 2009;

H.-C. Lim & Jang, 2006; Y. S. Park, 2011). Other notable issues include technological catch-up (S. J. Chang et al., 2006; Hobday, Rush, & Bessant, 2004; K.-H. Park & Lee, 2006; Shiu, Wong, & Hu, 2014; J. Yoon, 2015), crony capitalism (Ha & Lee, 2007), the role played by the military (Glassman & Choi, 2014; Maman, 2002), and the impact of economic and financial crises (Hundt, 2014; Y. S. Park, 2011; Shim, 2002; J.-H. Yoo & Moon, 1999).

The second largest community has 83 articles. It explores issues related to "corporate governance," such as family ownership and ownership structure (Baek, Kang, & Suh Park, 2004; Byun, Hwang, & Lee, 2011; J. Chang & Shin, 2007; S. J. Chang, 2003; Joh, 2003; J.-K. Kang, Lee, & Na, 2010; W. Kim, Lim, & Sung, 2007; U. Lim & Kim, 2005), internal capital markets, tunneling, and propping (Almeida et al., 2015; G. S. Bae et al., 2008; K.-H. Bae et al., 2002; Baek et al., 2006; Y. K. Choi, Han, & Kwon, 2019; Joh, 2003; R. Kim, 2016; S. Lee et al., 2009; H.-H. Shin & Park, 1999), agency theory and the role of outside directors (Chizema & Kim, 2010; J. J. Choi, Park, & Yoo, 2007; B. Kim & Lee, 2003; Min, 2013; Min & Verhoeven, 2013; J. Y. Shin et al., 2018; T. Yoo & Rhee, 2013), and corporate social responsibility (B. B. Choi, Lee, & Park, 2013; D. Choi, Choi, Choi, & Chung, 2020; Y. K. Choi et al., 2019; Chun & Shin, 2018; Jang, Ko, Chung, & Woo, 2019; A. Kim & Lee, 2018; B. Yoon, Lee, & Byun, 2018).

With 56 articles, "business groups in emerging markets" is the third largest community. Notable topics include the business environment in emerging markets (Khanna & Rivkin, 2001; Hicheon Kim et al., 2004; Mahmood & Mitchell, 2004; Rowley & Bae, 2004), technological innovation (Y. Kim & Lui, 2015; C.-Y. Lee et al., 2017; Mahmood & Mitchell, 2004), internal transactions (S. J. Chang & Hong, 2000; Gaur, Pattnaik, Singh, & Lee, 2019), networks, international expansion, and multinational firms (Guillen, 2002; S. W. Jeong, 2016; S. W. Jeong, Jin, & Jung, 2019; Hicheon Kim, Kim, & Hoskisson, 2010; Y. Kim & Lui, 2015; J. Y. Lee, Ryu, & Kang, 2014; Y. Park, Yul Lee, & Hong, 2011; Rugman

& Oh, 2008). The study by Khanna and Rivkin (2001), the highest cited article in the entire bibliographic coupling network, is an international comparative study of business groups based on Korean data.

The two smallest communities are "learning and human capital" (13 articles) and "financial liberalization" (7 articles). The former focuses on learning organizations, human capital, and leadership (Joo & Lee, 2017; D. H. Lim & Morris, 2006; J. H. Song, Joo, & Chermack, 2009), while the latter is concerned with investment, capital structure, and financial constraints (Hyesung Kim, Heshmati, & Aoun, 2006; Koo & Maeng, 2005; Krishnan & Moyer, 1997).

GRAPH 4 is a filtered version of GRAPH 3 and includes only the articles with the highest number of citations. The connections are the strongest between two pairs of articles: (K.-H. Bae et al., 2002) and (Baek et al., 2004), as well as (Baek et al., 2004) and (J.-B. Kim & Yi, 2006), with 14 articles in common in both cases. The next strongest link, with 11 articles in common, is connecting (Khanna & Rivkin, 2001) with (S. J. Chang & Hong, 2000). There are also two edges with a weight of 10, and those are the connection between (K.-H. Bae et al., 2002) and (Joh, 2003), and the link between (Joh, 2003) and (Baek et al., 2004). These high-weight edges between influential articles are predominantly linking nodes in the corporate governance community.

Compared to other articles in GRAPH 4, the study by Khanna and Rivkin (2001) has the most citations (930) and normalized citations (10.1). The articles with the highest normalized citation count in GRAPH 3 are two of the more recent studies. The first, with 111 citations and 10 normalized citations, explores the impact of corporate social responsibility on the market value of Korean firms (B. Yoon et al., 2018). The second, with 10 citations and 12 normalized citations, investigates the technology holding companies and the research activities of Korean universities (Son, Chung, & Yoon, 2022).

GRAPH 3 Bibliographic coupling network, articles with 10+ citations (size: citation count, color: modularity class)

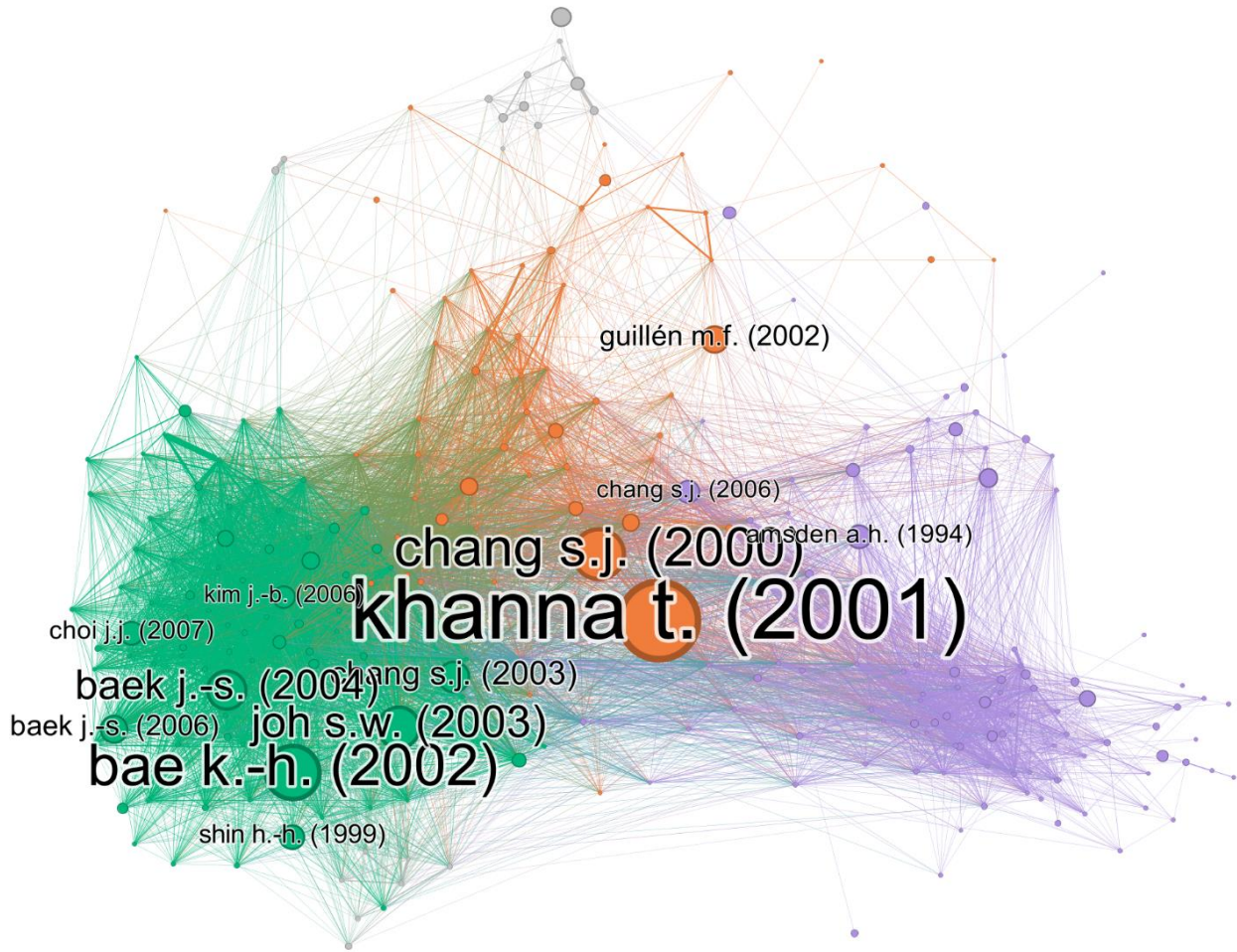

Source: Own elaboration based on Scopus (Elsevier, 2023).

GRAPH 4 Bibliographic coupling network, articles with 200+ citations (size: citation count, color: normalized citation count)

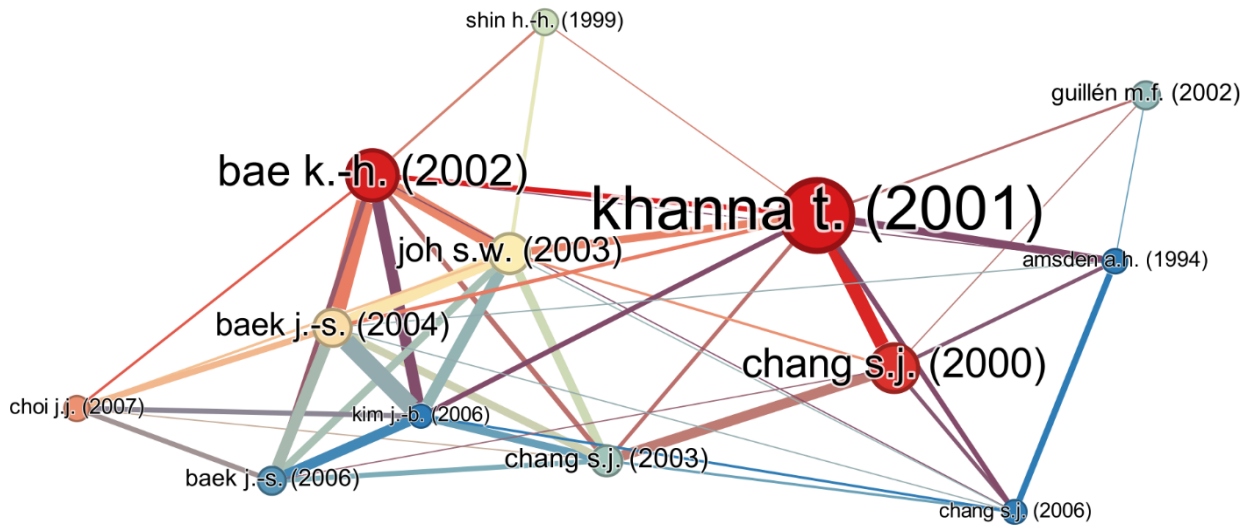

Source: Own elaboration based on Scopus (Elsevier, 2023).

TABLE 2 Bibliographic coupling network communities

| Number of articles | Community (color) | Notable themes | Highly cited articles | Citations | Normalized citations |
|---|---|---|---|---|---|
| 102 | Economic system (violet) | economic systems, developmental state, technological catch-up, crisis, capitalism, neoliberalism, free market, industrialization, military | (Amsden & Hikino, 1994) | 251 | 2.81 |
| | | | (S. J. Chang et al., 2006) | 238 | 2.88 |
| | | | (Hobday et al., 2004) | 187 | 3.00 |
| | | | (Kyung-Sup, 1999) | 156 | 3.34 |
| | | | (K.-H. Park & Lee, 2006) | 127 | 1.54 |
| 83 | Corporate governance (green) | corporate governance, ownership structure, internal capital markets, tunneling, propping, family ownership, agency theory, outside directors, executive compensation, corporate social responsibility | (K.-H. Bae et al., 2002) | 628 | 10.00 |
| | | | (Joh, 2003) | 475 | 6.71 |
| | | | (Baek et al., 2004) | 436 | 6.99 |
| | | | (S. J. Chang, 2003) | 342 | 4.83 |
| | | | (Baek et al., 2006) | 292 | 3.54 |
| 56 | Business groups in emerging markets (orange) | emerging markets, multinational enterprises, business networks, internal transactions, international expansion, technological innovation | (Khanna & Rivkin, 2001) | 930 | 10.10 |
| | | | (S. J. Chang & Hong, 2000) | 586 | 9.70 |
| | | | (Guillen, 2002) | 289 | 4.60 |
| | | | (Hicheon Kim et al., 2010) | 176 | 9.44 |
| | | | (Mahmood & Mitchell, 2004) | 170 | 2.73 |
| 13 | Learning and human capital (gray) | learning organizations, human capital, leadership, work engagement, career satisfaction | (D. H. Lim & Morris, 2006) | 199 | 2.41 |
| | | | (J. H. Song, Joo, et al., 2009) | 123 | 5.08 |
| | | | (Joo & Lee, 2017) | 83 | 7.98 |
| | | | (Joo, Lim, & Kim, 2016) | 67 | 8.89 |
| | | | (J. H. Song, Kim, & Kolb, 2009) | 62 | 2.56 |
| 7 | Financial liberalization (gray) | financial liberalization, investment, capital structure, financial constraints, corporate debt | (Krishnan & Moyer, 1997) | 47 | 1.55 |
| | | | (Hyesung Kim et al., 2006) | 33 | 0.40 |
| | | | (Koo & Maeng, 2005) | 28 | 1.31 |
| | | | (J.-W. Lee, Lee, & Lee, 2000) | 27 | 0.45 |
| | | | (Laeven, 2002) | 25 | 0.40 |

Source: Own elaboration based on Scopus (Elsevier, 2023).

*Keyword co-occurrence network and main themes of research*

The analysis of keyword co-occurrence networks has become a popular method of knowledge mapping. Unlike reference-based methods, research hotspots in co-occurrence networks are identified by analyzing the relationships between keywords. In this article, the network is based on author keywords. As shown in GRAPH 5, nodes represent keywords and their size is determined by occurrence count (the number of articles that a keyword occurs in). Edges represent the number of times two keywords co-occur. Much like in GRAPH 3, colors are determined by modularity class (community). Keywords are included if they occur in at least 2 articles For better readability, the labels are only shown for keywords that occur in 20 or more articles.

According to the occurrence count, the most common keywords are "chaebol" (352 articles), "south korea" (150 articles), "corporate governance" (90 articles), "financial crisis" (44 articles), "corporate social responsibility" (44 articles), "ownership structure" (32 articles), "innovation" (32 articles), "family firms" (30 articles), "developmental state" (30 articles), and "neoliberalism" (25 articles). As expected, the most significant edges are between "chaebol" and "south korea" (90 common articles) and "chaebol" and "corporate governance" (66 common articles). Degree centrality is the number of links to other keywords. The most connected keywords are "chaebol" (228 links), "south korea" (171 links), "corporate governance" (101 links), "financial crisis" (71 links), and "corporate social responsibility" (65 links).

As detailed in TABLE 3, the knowledge map has 12 thematic communities (resolution = 0.60). The largest community, with 59 keywords, is concerned with corporate governance and internal markets of chaebol groups. "Chaebol," the dominant keyword of the entire network (352 occurrences), is a part of this cluster. The research of corporate governance in business groups is primarily interested in ownership concentration, including the controlling family's stake, and the use of internal markets and related-party transactions.

Circular shareholding exacerbates the problems related to control-ownership disparity, tunneling (including propping), and agency problems.

The second largest community (53 keywords) comprises keywords related to the developmental state period of Korea and its subsequent liberalization. The focus lies on the institutional change and deregulation that occurred during the transformation from a developmental state to an advanced market economy. This cluster also includes keywords related to the globalization of Korea and its ties to other Asian economies, such as China, Japan, and Taiwan.

The next two communities consist of keywords related to corporate social responsibility, ESG, management, and accounting (26 keywords), and innovation, knowledge, learning, and patents (24 keywords). In total, the four largest communities (161 nodes) constitute 64% of the entire network (250 nodes).

The financial crisis in East Asia (19 keywords) is related to such keywords as investment, government regulation, credit ratings, the stock market, and performance. Various characteristics of Korean firms (18 keywords) include cash holdings, culture, capital structure, and risk. The information asymmetries and forecasts cluster (17 keywords) incorporates keywords like group affiliation, analyst following, banking sector, and forecast accuracy. Another community describes small and medium enterprises and business networks (15 keywords). Finally, the four smallest communities are earnings quality and sustainability (8 keywords), multinational firms and foreign direct investments (6 keywords), labor unions (4 keywords), and electronics suppliers (2 keywords).

GRAPH 5 Keyword co-occurrence network, keywords with 2+ occurrence count (size: occurrence count, color: modularity class)

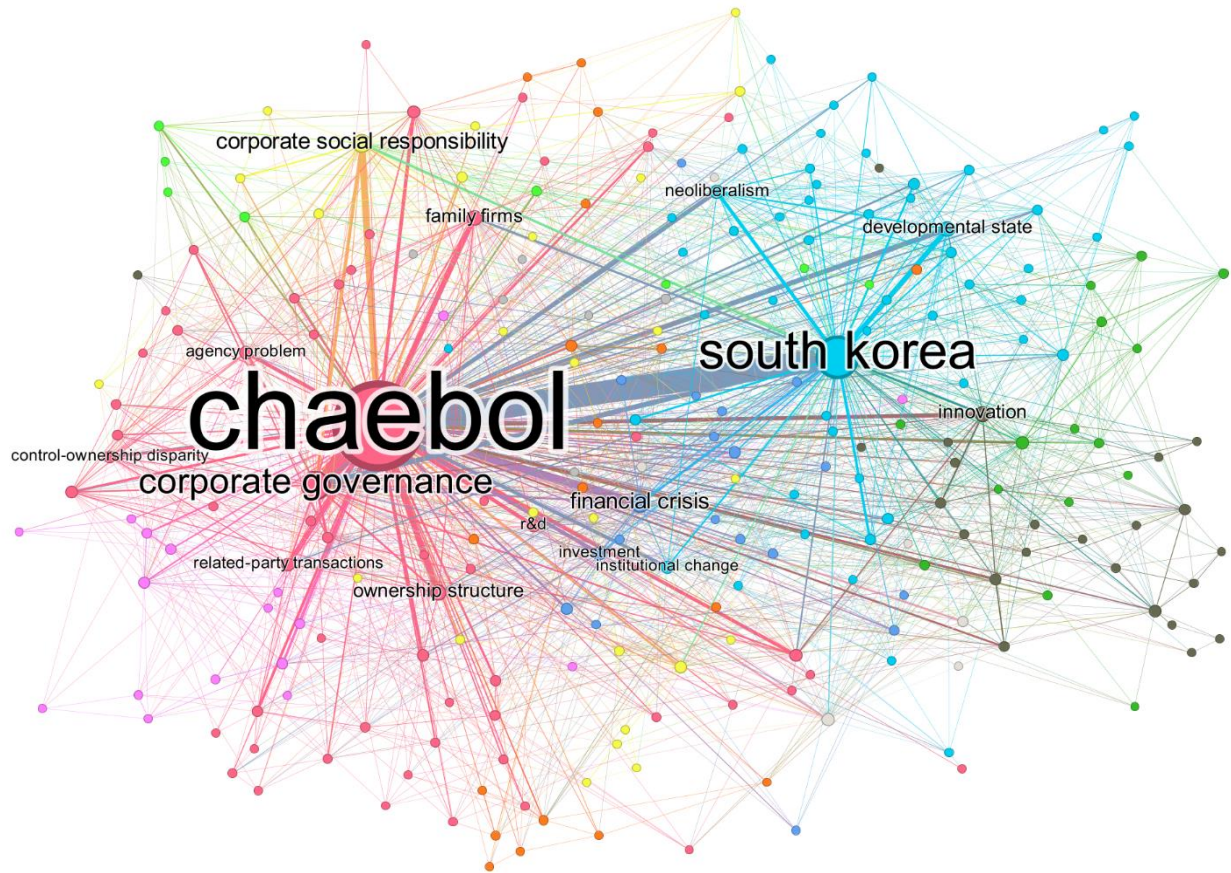

Source: Own elaboration based on Scopus (Elsevier, 2023).

TABLE 3 Keyword co-occurrence network communities

| Number of keywords | Community (color) | Top keywords | Occurrence count | Degree centrality |
|---|---|---|---|---|
| 59 | Corporate governance and internal markets of chaebol groups (red) | chaebol | 352 | 228 |
| | | corporate governance | 90 | 101 |
| | | ownership structure | 32 | 43 |
| | | family firms | 30 | 52 |
| | | r&d | 22 | 40 |
| | | related-party transactions | 22 | 32 |
| | | agency problem | 21 | 40 |
| | | control-ownership disparity | 20 | 36 |
| | | emerging market | 19 | 39 |
| | | firm value | 19 | 33 |
| | | controlling shareholder | 17 | 31 |
| | | internal capital market | 15 | 20 |
| | | institutional blockholding | 12 | 21 |
| | | tunneling | 12 | 23 |
| | | diversification | 11 | 17 |
| 52 | Developmental state and liberalization of Korea (blue) | south korea | 150 | 171 |
| | | developmental state | 30 | 51 |
| | | neoliberalism | 25 | 42 |
| | | institutional change | 20 | 41 |
| | | china | 13 | 34 |
| | | japan | 13 | 31 |
| | | technological capabilities | 13 | 21 |

| Number of keywords | Community (color) | Top keywords | Occurrence count | Degree centrality |
|---|---|---|---|---|
| | | globalization | 12 | 24 |
| | | taiwan | 11 | 21 |
| | | economic development | 10 | 19 |
| 26 | Corporate social responsibility and ESG (yellow) | corporate social responsibility | 44 | 65 |
| | | management | 17 | 35 |
| | | competition | 9 | 23 |
| | | environmentally sensitive industries | 8 | 17 |
| | | esg | 8 | 20 |
| | | accounting | 7 | 21 |
| | | ceo | 7 | 14 |
| | | discretionary accruals | 7 | 19 |
| 24 | Innovation and knowledge (dark gray) | innovation | 32 | 54 |
| | | human resource management | 17 | 24 |
| | | knowledge | 13 | 20 |
| | | learning | 12 | 32 |
| | | patent | 10 | 21 |
| | | subsidiary performance | 9 | 15 |
| 19 | Financial crisis in East Asia (dark blue) | financial crisis | 44 | 71 |
| | | investment | 22 | 35 |
| | | government regulation | 16 | 35 |
| | | credit ratings | 14 | 26 |
| | | korean stock market | 8 | 18 |
| | | performance | 8 | 21 |
| 18 | Characteristics of Korean firms (orange) | korean firms | 13 | 29 |
| | | cash holdings | 10 | 19 |
| | | culture | 9 | 17 |
| | | capital structure | 8 | 18 |
| | | risk | 7 | 18 |
| 17 | Information asymmetries and forecasts (pink) | information asymmetry | 14 | 30 |
| | | group affiliation | 13 | 24 |
| | | analyst following | 10 | 18 |
| | | banking sector | 9 | 26 |
| | | forecast accuracy | 9 | 13 |
| 15 | SMEs and business networks (dark green) | smes | 19 | 49 |
| | | business network | 8 | 20 |
| | | strategy | 8 | 21 |
| | | networks | 7 | 24 |
| | | resource-based view | 7 | 21 |
| | | social network analysis | 7 | 16 |
| 8 | Earnings quality and sustainability (green) | earnings quality | 9 | 21 |
| | | korean market | 7 | 19 |
| | | sustainable corporate governance | 7 | 16 |
| 6 | Multinational firms and FDI (light gray) | foreign direct investment | 18 | 41 |
| | | mnc | 12 | 27 |
| | | asia | 8 | 29 |
| 4 | Labor unions (gray) | labor union | 8 | 16 |
| 2 | Electronics suppliers (gray) | electronics industry | 2 | 4 |

Source: Own elaboration based on Scopus (Elsevier, 2023).

## 5. Implications

The bibliometric methodology provides tools for the large-scale systematization of existing knowledge. Quantification of research results has become increasingly important in academic research and evaluations. The bibliographic coupling network shows that scientific articles form distinct clusters around influential theories and studies. This can be used to trace the development of new theories, perhaps focusing on the more specialized aspects. Finally, the

keyword co-occurrence network maps the relationships between theoretical concepts, and, together with the lists of raw keywords and keyword aggregations provided in the associated dataset, can simplify the data collection process for future reviews and empirical studies of business groups.

International research collaboration is of paramount importance for research and development in both academic and corporate contexts. The list of prolific academic institutions should help to facilitate finding potential partners for university-industry collaborations. Business groups in Korea have traditionally been perceived as exploiters of their overwhelming market power. The fact that corporate social responsibility has gained a degree of prominence might suggest that systemic change had already occurred or will likely occur in the future. Similarly, the presence of foreign capital can alter the corporate governance of Korean firms. Whether these changes are positive (because they induce a shift to an incentive-based management style) or negative (because they dilute unique competitive advantages) is a different question entirely. Having said that, the strong internal markets of the chaebol remain one of their defining characteristics, despite international pressure and the government's attempts to regulate them. It does seem unlikely that Korean companies such as Samsung would have been able to successfully compete with American or Chinese multinationals had they not internalized so much of the market structures.

## 6. Conclusions

This article attempted to review and systematize the existing academic research on chaebol groups and their role in the economy of Korea. Business groups significantly contributed to the rapid economic growth during the developmental state period. However, due to their overwhelming market position, there have been numerous cases of corruption, cronyism, and state capture. Modern large business groups in Korea are heavily regulated and subject to rigorous disclosure requirements. Currently, the five largest chaebol are Samsung, Hyundai Motor, SK, LG, and Lotte. Ultimately, the economy of South Korea would not be in the strong position it is today without the chaebol.

RQ1 is answered with a co-authorship network which showed that the three most influential connections are between Korea and its scientific collaborations with the United States (119 articles), the United Kingdom (19 articles), and China (10 articles). Prolific journals include Sustainability (40 articles), Pacific-Basin Finance Journal (24 articles), Asia Pacific Business Review (19 articles), and Asia-Pacific Journal of Financial Studies (19 articles). Accordingly, Korean universities, such as Seoul National University (63 articles), Korea University (52 articles), Korea Advanced Institute of Science and Technology (44 articles), Yonsei University (41 articles), and Chung-Ang University (33 articles), play a key role in the research of business groups.

With regards to RQ2, bibliographic coupling identified five network communities of articles: "economic system" (102 articles), "corporate governance" (83 articles), "business groups in emerging markets" (56 articles), "learning and human capital" (13 articles), and "financial liberalization" (7 articles). The answer to RQ3 is revealed using a keyword co-occurrence network. There are 12 thematic communities of keywords, including "corporate governance and internal markets of chaebol groups" (59 keywords), "developmental state and liberalization of Korea" (52 keywords), "corporate social responsibility and ESG" (26 keywords), "innovation and knowledge" (24 keywords), and "financial crisis in East Asia" (19 keywords). Based on the occurrence count, the top keywords are "chaebol," "south korea," "corporate governance," "financial crisis," and "corporate social responsibility."

## References

Abbasi, A., Hossain, L., & Leydesdorff, L. (2012). Betweenness centrality as a driver of preferential attachment in the evolution of research collaboration networks. *Journal of Informetrics*, *6*(3), 403–412. https://doi.org/10.1016/j.joi.2012.01.002

Albrecht, C., Turnbull, C., Zhang, Y., & Skousen, C. J. (2010). The relationship between South Korean chaebols and fraud. *Management Research Review*, *33*(3), 257–268. https://doi.org/10.1108/01409171011030408

Al-Fadhat, F., & Choi, J.-W. (2023). Insights from the 2022 South Korean presidential election: Polarisation, fractured politics, inequality, and constraints on power. *Journal of Contemporary Asia*, 1–13. https://doi.org/10.1080/00472336.2023.2164937

Almeida, H., Kim, C.-S., & Kim, H. B. (2015). Internal capital markets in business groups: Evidence from the Asian financial crisis. *The Journal of Finance*, *70*(6), 2539–2586. https://doi.org/10.1111/jofi.12309

Almeida, H., Park, S. Y., Subrahmanyam, M. G., & Wolfenzon, D. (2011). The structure and formation of business groups: Evidence from Korean chaebols. *Journal of Financial Economics*, *99*(2), 447–475. https://doi.org/10.1016/j.jfineco.2010.08.017

Amsden, A. H., & Hikino, T. (1994). Project execution capability, organizational know-how and conglomerate corporate growth in late industrialization. *Industrial and Corporate Change*, *3*(1), 111–147. https://doi.org/10.1093/icc/3.1.111

Aoki, M., Patrick, H., & Sheard, P. (1994). The Japanese main bank system: An introductory overview. In M. Aoki & H. Patrick (Eds.), *The Japanese main bank system: Its relevance for developing and transforming economies* (pp. 1–50). Oxford University Press. https://doi.org/10.1093/0198288999.003.0001

Bae, G. S., Cheon, Y. S., & Kang, J.-K. (2008). Intragroup propping: Evidence from the stock-price effects of earnings announcements by Korean business groups. *Review of Financial Studies*, *21*(5), 2015–2060. https://doi.org/10.1093/rfs/hhn055

Bae, K.-H., Kang, J.-K., & Kim, J.-M. (2002). Tunneling or value added? Evidence from mergers by Korean business groups. *The Journal of Finance*, *57*(6), 2695–2740. https://doi.org/10.1111/1540-6261.00510

Baek, J.-S., Kang, J.-K., & Lee, I. (2006). Business groups and tunneling: Evidence from private securities offerings by Korean chaebols. *The Journal of Finance*, *61*(5), 2415–2449. https://doi.org/10.1111/j.1540-6261.2006.01062.x

Baek, J.-S., Kang, J.-K., & Suh Park, K. (2004). Corporate governance and firm value: Evidence from the Korean financial crisis. *Journal of Financial Economics*, *71*(2), 265–313. https://doi.org/10.1016/S0304-405X(03)00167-3

Barabási, A. L., Jeong, H., Néda, Z., Ravasz, E., Schubert, A., & Vicsek, T. (2002). Evolution of the social network of scientific collaborations. *Physica A: Statistical Mechanics and Its Applications*, *311*(3–4), 590–614. https://doi.org/10.1016/S0378-4371(02)00736-7

Bastian, M., Heymann, S., & Jacomy, M. (2009). Gephi: An open source software for exploring and manipulating networks. *Proceedings of the International AAAI Conference on Web and Social Media*, *3*(1), 361–362. https://doi.org/10.1609/icwsm.v3i1.13937

Beaver, D. deB., & Rosen, R. (1978). Studies in scientific collaboration: Part I. The professional origins of scientific co-authorship. *Scientometrics*, *1*(1), 65–84. https://doi.org/10.1007/BF02016840

Blondel, V. D., Guillaume, J.-L., Lambiotte, R., & Lefebvre, E. (2008). Fast unfolding of communities in large networks. *Journal of Statistical Mechanics: Theory and Experiment*, *2008*(10), P10008. https://doi.org/10.1088/1742-5468/2008/10/P10008

Buckley, P. J. (2004). Asian network firms: An analytical framework. *Asia Pacific Business Review*, *10*(3–4), 254–271. https://doi.org/10.1080/1360238042000264351

Byun, H.-Y., Choi, S., Hwang, L.-S., & Kim, R. G. (2013). Business group affiliation, ownership structure, and the cost of debt. *Journal of Corporate Finance*, *23*, 311–331. https://doi.org/10.1016/j.jcorpfin.2013.09.003

Byun, H.-Y., Hwang, L.-S., & Lee, W.-J. (2011). How does ownership concentration exacerbate information asymmetry among equity investors? *Pacific-Basin Finance Journal*, *19*(5), 511–534. https://doi.org/10.1016/j.pacfin.2011.06.002

Campbell, T. L., & Keys, P. Y. (2002). Corporate governance in South Korea: The chaebol experience. *Journal of Corporate Finance*, *8*(4), 373–391. https://doi.org/10.1016/S0929-1199(01)00049-9

Chang, J., Cho, Y. J., & Shin, H.-H. (2007). The change in corporate transparency of Korean firms after the Asian financial crisis: An analysis using analysts' forecast data. *Corporate Governance: An International Review*, *15*(6), 1144–1167. https://doi.org/10.1111/j.1467-8683.2007.00637.x

Chang, J., & Shin, H.-H. (2007). Family ownership and performance in Korean conglomerates. *Pacific-Basin Finance Journal*, *15*(4), 329–352. https://doi.org/10.1016/j.pacfin.2006.07.004

Chang, S. J. (2003). Ownership structure, expropriation, and performance of group-affiliated companies in Korea. *Academy of Management Journal*, *46*(2), 238–253. https://doi.org/10.2307/30040617

Chang, S. J., Chung, C.-N., & Mahmood, I. P. (2006). When and how does business group affiliation promote firm innovation? A tale of two emerging economies. *Organization Science*, *17*(5), 637–656. https://doi.org/10.1287/orsc.1060.0202

Chang, S. J., & Hong, J. (2000). Economic performance of group-affiliated companies in Korea: Intragroup-resource sharing and internal business transactions. *Academy of Management Journal*, *43*(3), 429–448. https://doi.org/10.2307/1556403

Chen, K., Zhang, Y., & Fu, X. (2019). International research collaboration: An emerging domain of innovation studies? *Research Policy*, *48*(1), 149–168. https://doi.org/10.1016/j.respol.2018.08.005

Cherry, J. (2003). The 'big deals' and Hynix semiconductor: State–business relations in post-crisis Korea. *Asia Pacific Business Review*, *10*(2), 178–198. https://doi.org/10.1080/13602380410001677209

Chizema, A., & Kim, J. (2010). Outside directors on Korean boards: Governance and institutions. *Journal of Management Studies*, *47*(1), 109–129. https://doi.org/10.1111/j.1467-6486.2009.00868.x

Cho, Y.-H., Yu, G.-C., Joo, M.-K., & Rowley, C. (2014). Changing corporate culture over time in South Korea. *Asia Pacific Business Review*, *20*(1), 9–17. https://doi.org/10.1080/13602381.2012.755321

Choe, H., & Lee, D. H. (2017). The structure and change of the research collaboration network in Korea (2000–2011): Network analysis of joint patents. *Scientometrics*, *111*(2), 917–939. https://doi.org/10.1007/s11192-017-2321-2

Choe, S., & Pattnaik, C. (2007). The transformation of Korean business groups after the Asian crisis. *Journal of Contemporary Asia*, *37*(2), 232–255. https://doi.org/10.1080/00472330701254062

Choi, B. B., Lee, D., & Park, Y. (2013). Corporate social responsibility, corporate governance and earnings quality: Evidence from Korea: csr, corporate governance and earnings quality. *Corporate Governance: An International Review*, *21*(5), 447–467. https://doi.org/10.1111/corg.12033

Choi, D., Choi, P. M. S., Choi, J. H., & Chung, C. Y. (2020). Corporate governance and corporate social responsibility: Evidence from the role of the largest institutional blockholders in the Korean market. *Sustainability*, *12*(4), 1680. https://doi.org/10.3390/su12041680

Choi, J. J., Park, S. W., & Yoo, S. S. (2007). The value of outside directors: Evidence from corporate governance reform in Korea. *Journal of Financial and Quantitative Analysis*, *42*(4), 941–962. https://doi.org/10.1017/S0022109000003458

Choi, Y. K., Han, S. H., & Kwon, Y. (2019). CSR activities and internal capital markets: Evidence from Korean business groups. *Pacific-Basin Finance Journal*, *55*, 283–298. https://doi.org/10.1016/j.pacfin.2019.04.008

Choung, J.-Y., & Hwang, H.-R. (2013). The evolutionary patterns of knowledge production in Korea. *Scientometrics*, *94*(2), 629–650. https://doi.org/10.1007/s11192-012-0780-z

Chun, H.-M., & Shin, S.-Y. (2018). Does analyst coverage enhance firms' corporate social performance? Evidence from Korea. *Sustainability*, *10*(7), 2561. https://doi.org/10.3390/su10072561

Chung, C. Y., Kim, D., & Lee, J. (2020). Do institutional investors improve corporate governance quality? Evidence from the blockholdings of the Korean National Pension Service. *Global Economic Review*, *49*(4), 422–437. https://doi.org/10.1080/1226508X.2020.1798268

Chung, C. Y., Kim, H., & Wang, K. (2022). Do domestic or foreign institutional investors matter? The case of firm information asymmetry in Korea. *Pacific-Basin Finance Journal*, *72*, 101727. https://doi.org/10.1016/j.pacfin.2022.101727

Dalton, B., & Rama, M. dela. (2016). Understanding the rise and decline of shareholder activism in South Korea: The explanatory advantages of the theory of Modes of Exchange. *Asia Pacific Business Review*, *22*(3), 468–486. https://doi.org/10.1080/13602381.2015.1129768

Ducret, R., & Isakov, D. (2020). The Korea discount and chaebols. *Pacific-Basin Finance Journal*, *63*, 101396. https://doi.org/10.1016/j.pacfin.2020.101396

Elsevier. (2023). Scopus database. Retrieved February 2, 2023, from https://www.scopus.com

Fernandes, A. J., & Ferreira, J. J. (2022). Entrepreneurial ecosystems and networks: A literature review and research agenda. *Review of Managerial Science*, *16*(1), 189–247. https://doi.org/10.1007/s11846-020-00437-6

Fitzgerald, R., & Kang, J. W. (2022). Transforming Korean business? Foreign acquisition, governance and management after the 1997 Asian crisis. *Asia Pacific Business Review*, *28*(1), 111–129. https://doi.org/10.1080/13602381.2021.1972612

Fukuda, S. (2018). Impacts of Japan's negative interest rate policy on Asian financial markets. *Pacific Economic Review*, *23*(1), 67–79. https://doi.org/10.1111/1468-0106.12253

Garfield, E. (1955). Citation indexes for science: A new dimension in documentation through association of ideas. *Science*, *122*(3159), 108–111. https://doi.org/10.1126/science.122.3159.108

Garner, J. L., & Kim, W. Y. (2013). Are foreign investors really beneficial? Evidence from South Korea. *Pacific-Basin Finance Journal*, *25*, 62–84. https://doi.org/10.1016/j.pacfin.2013.08.003

Gaur, A. S., Pattnaik, C., Singh, D., & Lee, J. Y. (2019). Internalization advantage and subsidiary performance: The role of business group affiliation and host country characteristics. *Journal of International Business Studies*, *50*(8), 1253–1282. https://doi.org/10.1057/s41267-019-00236-6

Gazni, A., & Didegah, F. (2011). Investigating different types of research collaboration and citation impact: A case study of Harvard University's publications. *Scientometrics*, *87*(2), 251–265. https://doi.org/10.1007/s11192-011-0343-8

Glänzel, W. (2001). National characteristics in international scientific co-authorship relations. *Scientometrics*, *51*(1), 69–115. https://doi.org/10.1023/A:1010512628145

Glänzel, W., & Schubert, A. (2001). Double effort = Double impact? A critical view at international co-authorship in chemistry. *Scientometrics*, *50*(2), 199–214. https://doi.org/10.1023/A:1010561321723

Glänzel, W., & Schubert, A. (2005). Analysing scientific networks through co-authorship. In H. F. Moed, W. Glänzel, & U. Schmoch (Eds.), *Handbook of quantitative science and technology research* (pp. 257–276). Dordrecht: Kluwer Academic Publishers. https://doi.org/10.1007/1-4020-2755-9_12

Glassman, J., & Choi, Y.-J. (2014). The chaebol and the US military—Industrial complex: Cold War geopolitical economy and South Korean Industrialization. *Environment and Planning A: Economy and Space*, *46*(5), 1160–1180. https://doi.org/10.1068/a130025p

Guillen, M. F. (2002). Structural inertia, imitation, and foreign expansion: South Korean firms and business groups in China, 1987-95. *Academy of Management Journal*, *45*(3), 509–525. https://doi.org/10.2307/3069378

Ha, Y.-C., & Lee, W. H. (2007). The politics of economic reform in South Korea: Crony capitalism after ten years. *Asian Survey*, *47*(6), 894–914. https://doi.org/10.1525/as.2007.47.6.894

Haggard, S. (2018). *Developmental states*. Cambridge University Press. https://doi.org/10.1017/9781108552738

Haggard, S., Kim, B.-K., & Moon, C. (1991). The transition to export-led growth in South Korea: 1954–1966. *The Journal of Asian Studies*, *50*(4), 850–873. https://doi.org/10.2307/2058544

Haggard, S., & Mo, J. (2000). The political economy of the Korean financial crisis. *Review of International Political Economy*, *7*(2), 197–218. https://doi.org/10.1080/096922900346947

Haggard, S., & Moon, C. (1990). Institutions and economic policy: Theory and a Korean case study. *World Politics*, *42*(2), 210–237. https://doi.org/10.2307/2010464

Han, J.-S., & Lee, J.-W. (2020). Demographic change, human capital, and economic growth in Korea. *Japan and the World Economy*, *53*, 100984. https://doi.org/10.1016/j.japwor.2019.100984

Han, S. (2022). Elite polarization in South Korea: Evidence from a natural language processing model. *Journal of East Asian Studies*, *22*(1), 45–75. https://doi.org/10.1017/jea.2021.36

Hobday, M., Rush, H., & Bessant, J. (2004). Approaching the innovation frontier in Korea: The transition phase to leadership. *Research Policy*, *33*(10), 1433–1457. https://doi.org/10.1016/j.respol.2004.05.005

Holmberg, K., & Park, H. W. (2018). An altmetric investigation of the online visibility of South Korea-based scientific journals. *Scientometrics*, *117*(1), 603–613. https://doi.org/10.1007/s11192-018-2874-8

Hundt, D. (2014). Economic crisis in Korea and the degraded developmental state. *Australian Journal of International Affairs*, *68*(5), 499–514. https://doi.org/10.1080/10357718.2014.886667

Jang, S. S., Ko, H., Chung, Y., & Woo, C. (2019). CSR, social ties and firm performance. *Corporate Governance: The International Journal of Business in Society*, *19*(6), 1310–1323. https://doi.org/10.1108/CG-02-2019-0068

Jeong, K.-Y., & Masson, R. T. (1990). Market structure, entry, and performance in Korea. *The Review of Economics and Statistics*, *72*(3), 455. https://doi.org/10.2307/2109353

Jeong, S. W. (2016). Types of foreign networks and internationalization performance of Korean SMEs. *Multinational Business Review*, *24*(1), 47–61. https://doi.org/10.1108/MBR-08-2015-0039

Jeong, S. W., Jin, B. E., & Jung, S. (2019). The temporal effects of social and business networks on international performance of South Korean SMEs. *Asia Pacific Journal of Marketing and Logistics*, *31*(4), 1042–1057. https://doi.org/10.1108/APJML-08-2018-0326

Jiang, Y., Ritchie, B. W., & Benckendorff, P. (2019). Bibliometric visualisation: An application in tourism crisis and disaster management research. *Current Issues in Tourism*, *22*(16), 1925–1957. https://doi.org/10.1080/13683500.2017.1408574

Jin, Y. (2020). Exclusionary effects of internal transactions of large business groups. *Global Economic Review*, *49*(3), 251–272. https://doi.org/10.1080/1226508X.2020.1745085

Joe, D. Y., Oh, F. D., & Yoo, H. (2019). Foreign ownership and firm innovation: Evidence from Korea. *Global Economic Review*, *48*(3), 284–302. https://doi.org/10.1080/1226508X.2019.1632729

Joh, S. W. (2003). Corporate governance and firm profitability: Evidence from Korea before the economic crisis. *Journal of Financial Economics*, *68*(2), 287–322. https://doi.org/10.1016/S0304-405X(03)00068-0

Joo, B.-K., & Lee, I. (2017). Workplace happiness: Work engagement, career satisfaction, and subjective well-being. *Evidence-Based HRM: A Global Forum for Empirical Scholarship*, *5*(2), 206–221. https://doi.org/10.1108/EBHRM-04-2015-0011

Joo, B.-K., Lim, D. H., & Kim, S. (2016). Enhancing work engagement: The roles of psychological capital, authentic leadership, and work empowerment. *Leadership & Organization Development Journal*, *37*(8), 1117–1134. https://doi.org/10.1108/LODJ-01-2015-0005

Jun, I., Sheldon, P., & Rhee, J. (2010). Business groups and regulatory institutions: Korea's chaebols, cross-company shareholding and the East Asian crisis. *Asian Business & Management*, *9*(4), 499–523. https://doi.org/10.1057/abm.2010.26

Jun, I.-W., Kim, K.-I., & Rowley, C. (2019). Organizational culture and the tolerance of corruption: The case of South Korea. *Asia Pacific Business Review*, *25*(4), 534–553. https://doi.org/10.1080/13602381.2019.1589728

Jung, B., Lee, D., Rhee, S. G., & Shin, I. (2019). Business group affiliation, internal labor markets, external capital markets, and labor investment efficiency. *Asia-Pacific Journal of Financial Studies*, *48*(1), 65–97. https://doi.org/10.1111/ajfs.12245

Jung, J.-K., & Choi, J. Y. (2022). Choice and allocation characteristics of faculty time in Korea: Effects of tenure, research performance, and external shock. *Scientometrics*, *127*(5), 2847–2869. https://doi.org/10.1007/s11192-022-04320-x

Kang, H. C., Anderson, R. M., Eom, K. S., & Kang, S. K. (2017). Controlling shareholders' value, long-run firm value and short-term performance. *Journal of Corporate Finance*, *43*, 340–353. https://doi.org/10.1016/j.jcorpfin.2017.01.013

Kang, J.-K., Lee, I., & Na, H. S. (2010). Economic shock, owner-manager incentives, and corporate restructuring: Evidence from the financial crisis in Korea. *Journal of Corporate Finance*, *16*(3), 333–351. https://doi.org/10.1016/j.jcorpfin.2009.12.001

Kang, S., Chung, C. Y., & Kim, D.-S. (2019). The effect of institutional blockholders' short-termism on firm innovation: Evidence from the Korean market. *Pacific-Basin Finance Journal*, *57*, 101188. https://doi.org/10.1016/j.pacfin.2019.101188

Kang, S. J., & Lee, H. S. (2007). The determinants of location choice of South Korean FDI in China. *Japan and the World Economy*, *19*(4), 441–460. https://doi.org/10.1016/j.japwor.2006.06.004

Kato, T., Kim, W., & Lee, J. H. (2007). Executive compensation, firm performance, and Chaebols in Korea: Evidence from new panel data. *Pacific-Basin Finance Journal*, *15*(1), 36–55. https://doi.org/10.1016/j.pacfin.2006.03.004

Kessler, M. M. (1963a). An experimental study of bibliographic coupling between technical papers. *IEEE Transactions on Information Theory*, *9*(1), 49–51. https://doi.org/10.1109/TIT.1963.1057800

Kessler, M. M. (1963b). Bibliographic coupling between scientific papers. *American Documentation*, *14*(1), 10–25. https://doi.org/10.1002/asi.5090140103

Khanna, T., & Rivkin, J. W. (2001). Estimating the performance effects of business groups in emerging markets. *Strategic Management Journal*, *22*(1), 45–74. https://doi.org/10.1002/1097-0266(200101)22:1<45::AID-SMJ147>3.0.CO;2-F

Kim, A., & Lee, Y. (2018). Family firms and corporate social performance: Evidence from Korean firms. *Asia Pacific Business Review*, *24*(5), 693–713. https://doi.org/10.1080/13602381.2018.1473323

Kim, B., & Lee, I. (2003). Agency problems and performance of Korean companies during the Asian financial crisis: Chaebol vs. non-chaebol firms. *Pacific-Basin Finance Journal*, *11*(3), 327–348. https://doi.org/10.1016/S0927-538X(03)00027-1

Kim, B.-K., & Im, H.-B. (2001). 'Crony capitalism' in South Korea, Thailand and Taiwan: Myth and reality. *Journal of East Asian Studies*, *1*(1), 5–52. https://doi.org/10.1017/S1598240800000230

Kim, C., Park, J.-H., Kim, J., & Lee, Y. (2022). The shadow of a departing CEO: Outsider succession and strategic change in a business group. *Asia Pacific Business Review*, *28*(1), 65–86. https://doi.org/10.1080/13602381.2021.1968655

Kim, E. (2006). The impact of family ownership and capital structures on productivity performance of Korean manufacturing firms: Corporate governance and the "chaebol problem." *Journal of the Japanese and International Economies*, *20*(2), 209–233. https://doi.org/10.1016/j.jjie.2005.02.001

Kim, E. M., & Oh, J. (2012). Determinants of foreign aid: The case of South Korea. *Journal of East Asian Studies*, *12*(2), 251–274. https://doi.org/10.1017/S1598240800007852

Kim, Hicheon, Hoskisson, R. E., Tihanyi, L., & Hong, J. (2004). The evolution and restructuring of diversified business groups in emerging markets: The lessons from chaebols in Korea. *Asia Pacific Journal of Management*, *21*(1/2), 25–48. https://doi.org/10.1023/B:APJM.0000024076.86696.d5

Kim, Hicheon, Kim, H., & Hoskisson, R. E. (2010). Does market-oriented institutional change in an emerging economy make business-group-affiliated multinationals perform better? An institution-based view. *Journal of International Business Studies*, *41*(7), 1141–1160. https://doi.org/10.1057/jibs.2010.17

Kim, Hohyun, & Han, S. H. (2018). Compensation structure of family business groups. *Pacific-Basin Finance Journal*, *51*, 376–391. https://doi.org/10.1016/j.pacfin.2018.09.002

Kim, Hyesung, Heshmati, A., & Aoun, D. (2006). Dynamics of capital structure: The case of Korean listed manufacturing companies. *Asian Economic Journal*, *20*(3), 275–302. https://doi.org/10.1111/j.1467-8381.2006.00236.x

Kim, J.-B., & Yi, C. H. (2006). Ownership structure, business group affiliation, listing status, and earnings management: Evidence from Korea. *Contemporary Accounting Research*, *23*(2), 427–464. https://doi.org/10.1506/7T5B-72FV-MHJV-E697

Kim, Jiyoung. (2017). Corporate financial structure of South Korea after Asian financial crisis: The chaebol experience. *Journal of Economic Structures*, *6*(1), 24. https://doi.org/10.1186/s40008-017-0085-8

Kim, Joongi. (2008). A forensic study of Daewoo's corporate governance: Does responsibility for the meltdown solely lie with the chaebol and Korea? *Northwestern Journal of International Law & Business*, *28*(2), 273–340.

Kim, Jounghyeon. (2020). Determinants of corporate bankruptcy: Evidence from chaebol and non-chaebol firms in Korea. *Asian Economic Journal*, *34*(3), 275–300. https://doi.org/10.1111/asej.12218

Kim, M.-J. (2001). *A bibliometric analysis of physics publications in Korea, 1994-1998*.

Kim, R. (2016). Financial weakness and product market performance: Internal capital market evidence. *Journal of Financial and Quantitative Analysis*, *51*(1), 307–332. https://doi.org/10.1017/S0022109016000077

Kim, W., Lim, Y., & Sung, T. (2007). Group control motive as a determinant of ownership structure in business conglomerates. *Pacific-Basin Finance Journal*, *15*(3), 213–252. https://doi.org/10.1016/j.pacfin.2006.05.003

Kim, Y., & Lui, S. S. (2015). The impacts of external network and business group on innovation: Do the types of innovation matter? *Journal of Business Research*, *68*(9), 1964–1973. https://doi.org/10.1016/j.jbusres.2015.01.006

Kim, Y. T. (1999). Neoliberalism and the decline of the developmental state. *Journal of Contemporary Asia*, *29*(4), 441–461. https://doi.org/10.1080/00472339980000231

Kim, Y. T. (2005). DJnomics and the transformation of the developmental state. *Journal of Contemporary Asia*, *35*(4), 471–484. https://doi.org/10.1080/00472330580000271

Koo, J., & Maeng, K. (2005). The effect of financial liberalization on firms' investments in Korea. *Journal of Asian Economics*, *16*(2), 281–297. https://doi.org/10.1016/j.asieco.2005.02.003

Korea Fair Trade Commission. (2022). *Annual report*.

Kōsai, Y. (1989). The postwar Japanese economy, 1945-1973. In P. Duus (Ed.), *The Cambridge history of Japan volume 6: The twentieth century* (pp. 494–538). Cambridge University Press. https://doi.org/10.1017/CHOL9780521223577.011

Krishnan, V. S., & Moyer, R. C. (1997). Performance, capital structure and home country: An analysis of Asian corporations. *Global Finance Journal*, *8*(1), 129–143. https://doi.org/10.1016/S1044-0283(97)90010-7

Kwon, I., & Park, C.-Y. (2021). Corporate governance and the efficiency of government R&D grants. *Global Economic Review*, *50*(4), 293–309. https://doi.org/10.1080/1226508X.2021.1908156

Kwon, K.-S., Park, H. W., So, M., & Leydesdorff, L. (2012). Has globalization strengthened South Korea's national research system? National and international dynamics of the Triple Helix of scientific co-authorship relationships in South Korea. *Scientometrics*, *90*(1), 163–176. https://doi.org/10.1007/s11192-011-0512-9

Kyung-Sup, C. (1999). Compressed modernity and its discontents: South Korean society in transtion. *Economy and Society*, *28*(1), 30–55. https://doi.org/10.1080/03085149900000023

Laeven, L. (2002). Financial constraints on investments and credit policy in Korea. *Journal of Asian Economics*, *13*(2), 251–269. https://doi.org/10.1016/S1049-0078(02)00111-2

Lambiotte, R., Delvenne, J.-C., & Barahona, M. (2014). Random walks, Markov processes and the multiscale modular organization of complex networks. *IEEE Transactions on Network Science and Engineering*, *1*(2), 76–90. https://doi.org/10.1109/TNSE.2015.2391998

Lee, C.-Y., Lee, J.-H., & Gaur, A. S. (2017). Are large business groups conducive to industry innovation? The moderating role of technological appropriability. *Asia Pacific Journal of Management*, *34*(2), 313–337. https://doi.org/10.1007/s10490-016-9481-0

Lee, J. Y., Ryu, S., & Kang, J. (2014). Transnational HR network learning in Korean business groups and the performance of their subsidiaries. *The International Journal of Human Resource Management*, *25*(4), 588–608. https://doi.org/10.1080/09585192.2013.792860

Lee, J.-W., Lee, Y. S., & Lee, B.-S. (2000). The determination of corporate debt in Korea. *Asian Economic Journal*, *14*(4), 333–356. https://doi.org/10.1111/1467-8381.t01-1-00113

Lee, K. Y., Chung, C. Y., & Morscheck, J. (2020). Does geographic proximity matter in active monitoring? Evidence from institutional blockholder monitoring of corporate governance in the Korean market. *Global Economic Review*, *49*(2), 150–170. https://doi.org/10.1080/1226508X.2019.1699846

Lee, P.-C., & Su, H.-N. (2010). Investigating the structure of regional innovation system research through keyword co-occurrence and social network analysis. *Innovation*, *12*(1), 26–40. https://doi.org/10.5172/impp.12.1.26

Lee, S. J. (2008). The politics of chaebol reform in Korea: Social cleavage and new financial rules. *Journal of Contemporary Asia*, *38*(3), 439–452. https://doi.org/10.1080/00472330802078519

Lee, S., Park, K., & Shin, H.-H. (2009). Disappearing internal capital markets: Evidence from diversified business groups in Korea. *Journal of Banking & Finance*, *33*(2), 326–334. https://doi.org/10.1016/j.jbankfin.2008.08.004

Lee, S.-J., & Han, T. (2006). The demise of "Korea, Inc.": Paradigm shift in Korea's developmental state. *Journal of Contemporary Asia*, *36*(3), 305–324. https://doi.org/10.1080/00472330680000191

Li, Y., Rong, Y., Ahmad, U. M., Wang, X., Zuo, J., & Mao, G. (2021). A comprehensive review on green buildings research: Bibliometric analysis during 1998–2018. *Environmental Science and Pollution Research*, *28*(34), 46196–46214. https://doi.org/10.1007/s11356-021-12739-7

Lim, D. H., & Morris, M. L. (2006). Influence of trainee characteristics, instructional satisfaction, and organizational climate on perceived learning and training transfer. *Human Resource Development Quarterly*, *17*(1), 85–115. https://doi.org/10.1002/hrdq.1162

Lim, H. (2009). Democratization and the transformation process in East Asian developmental states: Financial reform in Korea and Taiwan. *Asian Perspective*, *33*(1), 75–110. https://doi.org/10.1353/apr.2009.0026

Lim, H.-C., & Jang, J.-H. (2006). Neo-liberalism in post-crisis South Korea: Social conditions and outcomes. *Journal of Contemporary Asia*, *36*(4), 442–463. https://doi.org/10.1080/00472330680000281

Lim, U., & Kim, C.-S. (2005). Determinants of ownership structure: An empirical study of the Korean conglomerates. *Pacific-Basin Finance Journal*, *13*(1), 1–28. https://doi.org/10.1016/j.pacfin.2003.11.001

Mahmood, I. P., & Mitchell, W. (2004). Two faces: Effects of business groups on innovation in emerging economies. *Management Science*, *50*(10), 1348–1365. https://doi.org/10.1287/mnsc.1040.0259

Maman, D. (2002). The emergence of business groups: Israel and South Korea compared. *Organization Studies*, *23*(5), 737–758. https://doi.org/10.1177/0170840602235003

Min, B.-S. (2013). Evaluation of board reforms: An examination of the appointment of outside directors. *Journal of the Japanese and International Economies*, *29*, 21–43. https://doi.org/10.1016/j.jjie.2013.05.001

Min, B.-S., & Verhoeven, P. (2013). Outsider board activity, ownership structure and firm value: Evidence from Korea: outsider board activity and firm value. *International Review of Finance*, *13*(2), 187–214. https://doi.org/10.1111/irfi.12004

Nam, H.-J., & An, Y. (2023). Interlocking network, banker interlocking director and cost of debt. *Global Economic Review*, *52*(1), 71–92. https://doi.org/10.1080/1226508X.2023.2171458

Newman, M. E. J. (2001). The structure of scientific collaboration networks. *Proceedings of the National Academy of Sciences*, *98*(2), 404–409. https://doi.org/10.1073/pnas.98.2.404

Oh, I., & Park, H.-J. (2001). Shooting at a moving target: Four theoretical problems in explaining the dynamics of the chaebol. *Asia Pacific Business Review*, *7*(4), 44–69. https://doi.org/10.1080/713999115

Oh, I., Yoon, S. S., & Kim, H. J. (2022). The end of rent sharing: Corporate governance reforms in South Korea. *Asia Pacific Business Review*, *28*(1), 16–37. https://doi.org/10.1080/13602381.2021.1961430

Oliveira, A., Carvalho, F., & Reis, N. R. (2022). Institutions and firms' performance: A bibliometric analysis and future research avenues. *Publications*, *10*(1), 8. https://doi.org/10.3390/publications10010008

Park, C.-K., Lee, C., & Jeon, J. Q. (2020). Centrality and corporate governance decisions of Korean chaebols: A social network approach. *Pacific-Basin Finance Journal*, *62*, 101390. https://doi.org/10.1016/j.pacfin.2020.101390

Park, H. W., Yoon, J., & Leydesdorff, L. (2016). The normalization of co-authorship networks in the bibliometric evaluation: The government stimulation programs of China and Korea. *Scientometrics*, *109*(2), 1017–1036. https://doi.org/10.1007/s11192-016-1978-2

Park, K.-H., & Lee, K. (2006). Linking the technological regime to the technological catch-up: Analyzing Korea and Taiwan using the US patent data. *Industrial and Corporate Change*, *15*(4), 715–753. https://doi.org/10.1093/icc/dtl016

Park, S. H. (2022). The tax models in Japan and Korea: Concepts and evidence from a comparative perspective. *Journal of East Asian Studies*, *22*(3), 359–389. https://doi.org/10.1017/jea.2022.13

Park, S., & Yuhn, K. (2012). Has the Korean chaebol model succeeded? *Journal of Economic Studies*, *39*(2), 260–274. https://doi.org/10.1108/01443581211222680

Park, Y. S. (2011). Revisiting the South Korean developmental state after the 1997 financial crisis. *Australian Journal of International Affairs*, *65*(5), 590–606. https://doi.org/10.1080/10357718.2011.607148

Park, Y., Yul Lee, J., & Hong, S. (2011). Effects of international entry-order strategies on foreign subsidiary exit: The case of Korean chaebols. *Management Decision*, *49*(9), 1471–1488. https://doi.org/10.1108/00251741111173943

Price, D. J. de S. (1965). Networks of scientific papers: The pattern of bibliographic references indicates the nature of the scientific research front. *Science*, *149*(3683), 510–515. https://doi.org/10.1126/science.149.3683.510

Rowley, C., & Bae, J. (2004). Big business in South Korea: The reconfiguration process. *Asia Pacific Business Review*, *10*(3–4), 302–323. https://doi.org/10.1080/1360238042000264388

Rugman, A. M., & Oh, C. H. (2008). Korea's multinationals in a regional world. *Journal of World Business*, *43*(1), 5–15. https://doi.org/10.1016/j.jwb.2007.10.003

Schwak, J. (2019). Dangerous liaisons? State-chaebol co-operation and the global privatisation of development. *Journal of Contemporary Asia*, *49*(1), 104–126. https://doi.org/10.1080/00472336.2018.1501806

Sheard, P. (1994). Main banks and the governance of financial distress. In M. Aoki & H. Patrick (Eds.), *The Japanese main bank system: Its relevance for developing and transforming economies* (pp. 188–230). Oxford University Press. https://doi.org/10.1093/0198288999.003.0006

Shim, D. (2002). South Korean media industry in the 1990s and the economic crisis. *Prometheus*, *20*(4), 337–350. https://doi.org/10.1080/0810902021000023336

Shin, H.-H., & Park, Y. S. (1999). Financing constraints and internal capital markets: Evidence from Korean `chaebols'. *Journal of Corporate Finance*, *5*(2), 169–191. https://doi.org/10.1016/S0929-1199(99)00002-4

Shin, J. Y., Hyun, J.-H., Oh, S., & Yang, H. (2018). The effects of politically connected outside directors on firm performance: Evidence from Korean chaebol firms. *Corporate Governance: An International Review*, *26*(1), 23–44. https://doi.org/10.1111/corg.12203

Shin, K., & Wang, Y. (2004). Trade integration and business cycle co-movements: The case of Korea with other Asian countries. *Japan and the World Economy*, *16*(2), 213–230. https://doi.org/10.1016/S0922-1425(03)00028-8

Shin, S. I., & Kim, L. (2020). Chaebol and the turn to services: The rise of a Korean service economy and the dynamics of self-employment and wage work. *Journal of Contemporary Asia*, *50*(3), 433–456. https://doi.org/10.1080/00472336.2019.1565130

Shiu, J.-W., Wong, C.-Y., & Hu, M.-C. (2014). The dynamic effect of knowledge capitals in the public research institute: Insights from patenting analysis of ITRI (Taiwan) and ETRI (Korea). *Scientometrics*, *98*(3), 2051–2068. https://doi.org/10.1007/s11192-013-1158-6

Small, H. (1973). Co-citation in the scientific literature: A new measure of the relationship between two documents. *Journal of the American Society for Information Science*, *24*(4), 265–269. https://doi.org/10.1002/asi.4630240406

Son, H., Chung, Y., & Yoon, S. (2022). How can university technology holding companies bridge the Valley of Death? Evidence from Korea. *Technovation*, *109*, 102158. https://doi.org/10.1016/j.technovation.2020.102158

Song, H. K. (2003). The birth of a welfare state in Korea: The unfinished symphony of democratization and globalization. *Journal of East Asian Studies*, *3*(3), 405–432. https://doi.org/10.1017/S1598240800001582

Song, J. H., Joo, B.-K. (Brian), & Chermack, T. J. (2009). The Dimensions of Learning Organization Questionnaire (DLOQ): A validation study in a Korean context. *Human Resource Development Quarterly*, *20*(1), 43–64. https://doi.org/10.1002/hrdq.20007

Song, J. H., Kim, H. M., & Kolb, J. A. (2009). The effect of learning organization culture on the relationship between interpersonal trust and organizational commitment. *Human Resource Development Quarterly*, *20*(2), 147–167. https://doi.org/10.1002/hrdq.20013

Su, H.-N., & Lee, P.-C. (2010). Mapping knowledge structure by keyword co-occurrence: A first look at journal papers in Technology Foresight. *Scientometrics*, *85*(1), 65–79. https://doi.org/10.1007/s11192-010-0259-8

Szczech-Pietkiewicz, E., Radło, M.-J., & Tomeczek, A. F. (2023). Smart and sustainable city management in Asia and Europe: A bibliometric analysis. In R. Dygas & P. K. Biswas (Eds.), *Smart Cities in Europe and Asia* (1st ed., pp. 1–15). London: Routledge. https://doi.org/10.4324/9781003365174-1

Tomeczek, A. F. (2022). The evolution of Japanese keiretsu networks: A review and text network analysis of their perceptions in economics. *Japan and the World Economy*, *62*, 101132. https://doi.org/10.1016/j.japwor.2022.101132

Tomeczek, A. F. (2023). *The rise of the chaebol: A bibliometric analysis of business groups in South Korea (data and graphs)* [Data set]. Mendeley Data. https://doi.org/10.17632/3fgc6bzyjp.1

Uddin, S., Hossain, L., Abbasi, A., & Rasmussen, K. (2012). Trend and efficiency analysis of co-authorship network. *Scientometrics*, *90*(2), 687–699. https://doi.org/10.1007/s11192-011-0511-x

van Eck, N. J., & Waltman, L. (2010). Software survey: VOSviewer, a computer program for bibliometric mapping. *Scientometrics*, *84*(2), 523–538. https://doi.org/10.1007/s11192-009-0146-3

van Eck, N. J., & Waltman, L. (2022). *VOSviewer manual*. Retrieved from https://www.vosviewer.com/documentation/Manual_VOSviewer_1.6.18.pdf

Velez-Estevez, A., García-Sánchez, P., Moral-Munoz, J. A., & Cobo, M. J. (2022). Why do papers from international collaborations get more citations? A bibliometric analysis of Library and Information Science papers. *Scientometrics*, *127*(12), 7517–7555. https://doi.org/10.1007/s11192-022-04486-4

Wang, D., Huangfu, Y., Dong, Z., & Dong, Y. (2022). Research hotspots and evolution trends of carbon neutrality—Visual analysis of bibliometrics based on citespace. *Sustainability*, *14*(3), 1078. https://doi.org/10.3390/su14031078

Xu, X., Hou, G., & Wang, J. (2022). Research on digital transformation based on complex systems: Visualization of knowledge maps and construction of a theoretical framework. *Sustainability*, *14*(5), 2683. https://doi.org/10.3390/su14052683

Yadav, N., Kumar, R., & Malik, A. (2022). Global developments in coopetition research: A bibliometric analysis of research articles published between 2010 and 2020. *Journal of Business Research*, *145*, 495–508. https://doi.org/10.1016/j.jbusres.2022.03.005

Yan, E., & Ding, Y. (2012). Scholarly network similarities: How bibliographic coupling networks, citation networks, cocitation networks, topical networks, coauthorship networks, and coword networks relate to each other. *Journal of the American Society for Information Science and Technology*, *63*(7), 1313–1326. https://doi.org/10.1002/asi.22680

Yan, E., Ding, Y., & Zhu, Q. (2010). Mapping library and information science in China: A coauthorship network analysis. *Scientometrics*, *83*(1), 115–131. https://doi.org/10.1007/s11192-009-0027-9

Yoo, J.-H., & Moon, C. W. (1999). Korean financial crisis during 1997–1998. Causes and challenges. *Journal of Asian Economics*, *10*(2), 263–277. https://doi.org/10.1016/S1049-0078(99)00018-4

Yoo, S., & Lee, S. M. (1987). Management style and practice of Korean chaebols. *California Management Review*, *29*(4), 95–110. https://doi.org/10.2307/41162133

Yoo, T., & Koh, Y. (2022). Remains on the board: Outside directors' behaviour and their survival chance in Korean firms. *Asia Pacific Business Review*, *28*(1), 87–110. https://doi.org/10.1080/13602381.2021.1956814

Yoo, T., & Rhee, M. (2013). Agency theory and the context for R&D investment: Evidence from Korea. *Asian Business & Management*, *12*(2), 227–252. https://doi.org/10.1057/abm.2013.2

Yoon, B., Lee, J., & Byun, R. (2018). Does ESG performance enhance firm value? Evidence from Korea. *Sustainability*, *10*(10), 3635. https://doi.org/10.3390/su10103635

Yoon, J. (2015). The evolution of South Korea's innovation system: Moving towards the triple helix model? *Scientometrics*, *104*(1), 265–293. https://doi.org/10.1007/s11192-015-1541-6

You, J. (2020). The changing dynamics of state–business relations and the politics of reform and capture in South Korea. *Review of International Political Economy*, *28*(1), 81–102. https://doi.org/10.1080/09692290.2020.1724176

You, J., & Park, Y. M. (2017). The legacies of state corporatism in Korea: Regulatory capture in the Sewol ferry tragedy. *Journal of East Asian Studies*, *17*(1), 95–118. https://doi.org/10.1017/jea.2016.32

Zhang, R., & Yuan, J. (2022). Enhanced author bibliographic coupling analysis using semantic and syntactic citation information. *Scientometrics*, *127*(12), 7681–7706. https://doi.org/10.1007/s11192-022-04333-6

Zhang, X., Xie, Q., Song, C., & Song, M. (2022). Mining the evolutionary process of knowledge through multiple relationships between keywords. *Scientometrics*, *127*(4), 2023–2053. https://doi.org/10.1007/s11192-022-04272-2